# Analyzing China's Consumer Price Index Comparatively with that of United States

Zhenzhong Wang[a], Song Xi Chen[b], and Yundong Tu[b,*]

[a]Department of Statistics, Iowa State University, Ames, IA 50011, USA
[b]Department of Business Statistics and Econometrics, Guanghua School of Management, Peking University, China
*Corresponding author: Yundong Tu, yundong.tu@gsm.pku.edu.cn.

November 30, 2018

### Abstract

This paper provides a thorough analysis on the dynamic structures and predictability of China's Consumer Price Index (CPI-CN), with a comparison to those of the United States. Despite the differences in the two leading economies, both series can be well modeled by a class of S-ARIMAX models. The CPI-CN series possess regular patterns of dynamics with stable annual cycles and strong Spring Festival effects, with fitting and forecasting errors largely comparable to their US counterparts. Finally, for the CPI-CN, the diffusion index (DI) approach offers improved predictions than the S-ARIMAX models.

# 1 Introduction

China has become the second largest economy in the world since 2011, and is increasingly an engine of the global economic growth. As China gains more prominence and influence in the world economy, the need to study the structures of the macroeconomic variables of China becomes more urgent. Among the macroeconomic variables, the consumer price index (CPI) is a key monthly index for the overall consumer price movement, and its level is a particularly important indicator for the monetary authority to maintain stable prices while pursing other economic objectives.

Despite the importance the Chinese CPI (CPI-CN), the study of its dynamics and structure is rather limited. Chow (1987), Chow (2007) and Chow and Wang (2010) represent pioneer studies on the Chinese price index. See also Zhang (2013) and Wang et al. (2016). One indication of the status is the fact that the official CPI in China is not seasonally adjusted due to a lack of understanding on both the seasonal and dynamic structures of the index. This means that the CPI data are likely to present features such as seasonal variations and holiday effects, which make the index series neither comparable from month to month for a given year, nor across years for a given month. The latter is due to the fact that important Chinese holidays, such as the Spring Festival (SF), are based on the lunar calendar while the prices are measured according to solar calendar. Consequently, any analysis based on the CPI-CN series without accounting for these features is likely to be misleading.[1]

There have been concerns on the potential data interference by the local and provincial authorities on China's macroeconomic data, which may contribute to the generally lack of studies on Chinese macroeconomic data. However, the CPI-CN has not been a performance indicator in various levels of the government, unlike the Gross Domestic Product (GDP) which had been the key performance measure for various level of administrators from provincial to city and county levels down. Hence, there is little incentive for manipulation and interference by the local authorities. In addition, the price index has been composed centrally by the National Bureau of Statistics (NBS) based on consumer prices collected by its City Survey Teams, which are directly funded by the NBS bypassing the local authorities.[2]

---

[1] For example, when the January 2015 CPI-CN rate of 0.8% (annually) was released, there was a wide spread interpretation of deflation by economic commentators, since it was almost 50% lower than the previous month's rate of 1.5% . However, the reason behind the dramatic decline was mainly the SF effect, as the date of the SF migrated from January 31 in 2014 to February 18th in 2015.

This paper aims at providing a thorough analysis of the CPI-CN on its dynamic structures and its predictability, comparatively to those of the United States. The study has the following objectives. First, we analyse the basic dynamic structures of CPI-CN by the conventional linear time series approach within the family of seasonal ARIMA models (S-ARIMA), augmented with the treatment of the holiday effect, especially the SF effect, and the outliers caused by major one-off events. Our study reveals that the CPI-CN series possess regular patterns of dynamics with stable annual cycles, and regular and yet strong SF effects. The linear time series models produce reasonable fitting performance, where the large errors can be explained by one-off natural and economic events (outliers) that trigger the large price movements. This analysis pins down a group of eight models that provide the best fits to the data and the best forecasting performance among a total of 11907 [3] models.

The second objective is to compare the predictability of the CPI-CN by the top eight linear time series models with that rendered by the top eight models for the US CPI (CPI-US) over the period of 2002 to 2016. To gain comparative perspectives on the fitting and forecasting errors on the CPI-CN series, a similar analysis on the CPI-US data with the linear time series models is performed which indicates that the fitting and the forecasting errors for CPI-US are largely comparable to those for CPI-CN.

The third objective is to see if the forecasting performance of the CPI-CN and CPI-

---

[2]A reason for this centrally administered CPI composition is the lesson learned from the political turmoil in 1989. There has been consensus among the upper circle of the government that the double digit inflation from the mid-1980s was a key factor that contributed to a wide spread discontent in the population prior to the event in 1989. Hence, the central government wants to know the genuine price index rather than the one which might have been compromised with local interferences. These two aspects suggest that the CPI-CN figures are of fairly good quality, as revealed by this study when we look at the finer dynamic structures of the price series.

[3]There are 81 S-ARIMA models, and each model has 147 choices of the Spring Festival effects, which lead to a total of 11907 models.

US can be improved by incorporating other macro-economic variables via the diffusion index (DI) approach proposed by Stock and Watson (2002b). Our analysis shows that for the CPI-CN, the DI approach offers improved forecasting performance than that of the univariate linear time series models when the forecasting horizon is three months or shorter; while for CPI-US, the DI approach behaves worse than the univariate linear time series model.

The structure of the paper is organized as follows. Section 2 introduces the CPI-CN and CPI-US series and their compositions. Section 3 outlines the modeling strategy using univariate Seasonal ARIMAX (S-ARIMAX) and the results obtained for CPI-CN. Section 4 applies the DI approach for the CPI-CN series to provide alternative approach for forecasting the price series. Section 5 reports the results for CPI-US which consist of the linear time series modeling and forecasting with the DI. Section 6 concludes with remarks. Additional results which are supplement to the main analysis are provided in the Supplementary Material.

## 2 CPI Data

The CPI series of China and the US, used in our study, range from January 2002 to November 2016. This choice of data range is made for two reasons. First, China joined the World Trade Organization in December 2001, since then its economy and the measurement of economic activities has become more compatible to the international norm. In particular, the CPI composition adopted in China after 2001 becomes quite similar to that in US. Second, the CPI series of the US also underwent a major revision in January 1999, when it began to adopt geometric means to measure the average price changes within most of the item categories (Bureau of Labor Statistics (BLS) 2007). As a result, the CPI series starting from 2002 offer the earliest opportunity for a meaningful comparison between the two countries' consumer price series.

In this paper, the CPI-U series (CPI for All Urban Consumers) is taken as the CPI-US, which is the headline price series composed by the BLS. China produces one national level CPI series (CPI-CN) for both urban and rural consumers, which is the most important CPI series in China. The CPI-U published by BLS has two versions: one with and the other without the seasonal adjustment. We choose the latter to be consistent with the fact that CPI-CN is not seasonally adjusted.

The compositions of CPI-CN and those of CPI-US are similar, with eight categories in each series. China's eight categories are Food; Alcohol and Tobacco; Clothing; Health Cares; Household Articles and Services; Transportation and Communication; Education, Culture and Recreation; Housing.[4] And those of the US include Food and Beverages; Apparel; Housing; Medical care; Transportation; Recreation; Education and Communication; Other Goods and Services. The two categories which are largely the same between the two countries are Housing and Apparel. However, the make-up of the other six categories are slightly different. China has a separate Tobacco and Alcohol sub-category before 2016, while the US has two sub-categories in Food and Beverage, and Other Goods and Services, respectively. China combines Recreation, Education and Culture Articles in one category while the US has a separate Recreation category, and merges Communication with Education. China's Household Facilities and Related Articles and Services largely fall to the sub-category of Other Goods and Services in the US index. The Personal Article sub-category in China's Medical Care and Personal Articles category would be in Other Goods and Services in the US index.

There are layers of weights which are used to produce price indices at higher level aggregates, including the sub-category level and the overall CPI. For both countries, the weights are obtained by consumer expenditure surveys. Specifically, the weights in CPI-US are obtained from data collected in the biennial Consumer Expenditure Survey (CES), while the weights in CPI-CN are based on Nationwide Household Expenditure Survey (NHES) conducted every five years.

Figure 1 presents the weights of the eight categories of CPI-CN (a) and CPI-US (b). As the official weights in the CPI-CN are not available, the weights shown in Figure 1 are estimates obtained by regressing CPI-CN on its eight categories based on the official CPI releases from years 2011 to 2014. To be consistent, Figure 1 (b) displays the regression estimates of the weights for the CPI-US using data in years 2013 and 2014, considering the biennial weight adjustment. The estimated weights for the CPI-US are very close to the published Relative Importance of the eight categories published by the BLS for the same period. Figure 1 (a) shows the weight composition for the CPI-CN, which shows Food takes the biggest weight at 31%, followed by Housing at 21%. Weights of the other six categories are all less than 12%. In contrast, in the CPI-US, Housing is the dominant category occupying 42% weight, followed by Transportation at around 18%. Food and Beverages is at 15.1%, much lower than its Chinese counterpart. Hence, CPI-CN is much dominated by Food and Housing, while CPI-US by Housing and Transportation, respectively.

---

[4]From January 2016, sub-categories of CPI-CN are re-structured as: Food and Tobacco; Clothing; Health Cares; Household Articles and Services; Transportation and Communication; Education, Culture and Recreation; Housing; Miscellaneous Goods and Services.

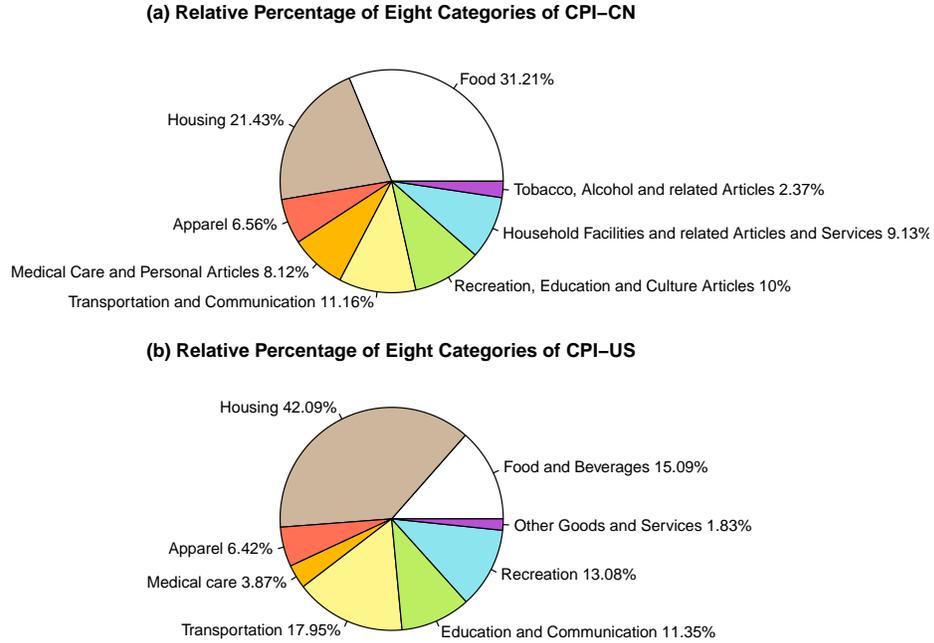

Figure 1: compositions of CPI-CN (a) and CPI-US (b).

Figure 2 (a) and Figure 2 (c) plot the Raw CPI series of China and US respectively. It appears that the last financial crisis had much bigger impact on the CPI-US than on the CPI-CN. This was not only because the Unite States was the epicenter of the crisis, but also has to do with the different CPI composition of the two countries. The CPI-US is very much housing dominant. And it was the bursting of US housing bubble that induced the financial crisis. In contrast, the CPI-CN is largely food dominant, whose demand and supply are much less affected by the crisis.

Figure 2 (b) shows the seasonality of CPI-CN extracted via the X-13ARIMA-SEATS

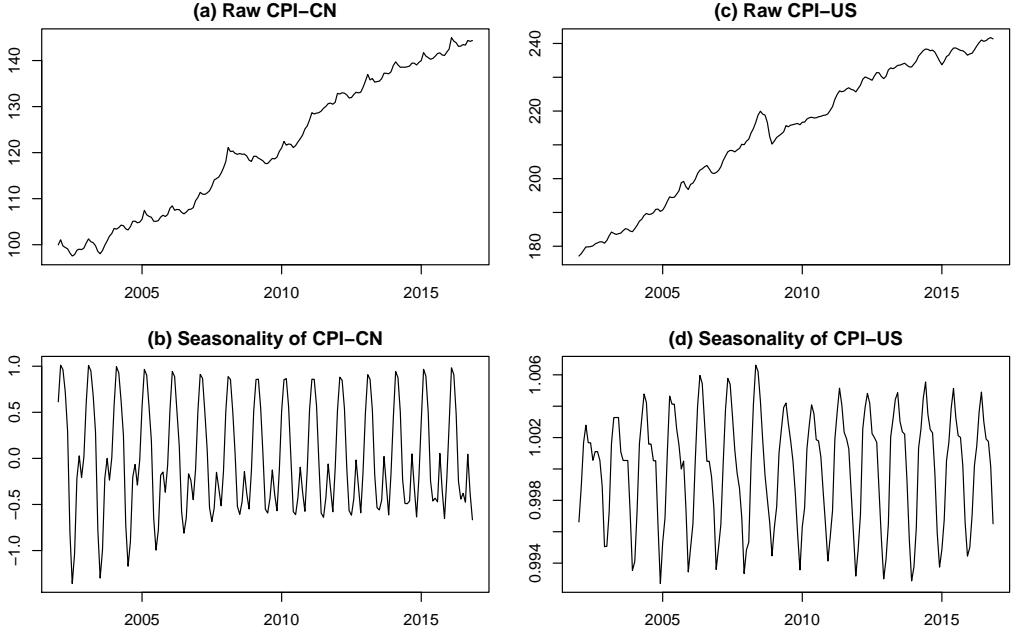

Figure 2: The Panel (a) and (b) plots the raw series of CPI-CN and its seasonality, respectively; the Panel (c) and (d) plots the raw series of CPI-US and its seasonality, respectively.

procedure (U.S. Census Bureau, 2017), which is the official seasonal adjustment procedure provided by the United States Census Bureau. It shows that the CPI-CN has a pronounced annual seasonality, which peaks in February and reaches its trough in July. The magnitude of the annual cycle was reduced gradually to a new level since 2007. And the seasonality of CPI-US shown in Figure 2 (d) peaks usually around May and gets its trough in December. As the CPI-CN are released without undergoing seasonal adjustment, the seasonality will be modeled in the next section.

Apart from the fact that China's CPI is not seasonally adjusted, the other major difference between CPI-CN and CPI-US is that CPI-CN is subject to lunar holiday effects. Some important Chinese holidays such as the Spring Festival, Duanwu and Mid-Autumn Day are based on the lunar calendar, and their dates can vary across two adjacent calendar months in two adjacent years. Consumption and production activities in China are strongly affected by these holidays, especially the Spring Festival. Specifically, the holidays increase consumers' demand for goods and services while the supply of the goods and service may be reduced, especially during the Spring Festival period, a time that most offices and factories are closed and the biggest annual migration back to the countryside occurs. As a result, there is often dramatic increase in prices, which causes large fluctuations in CPI-CN.

In contrast, the dates of US holidays tend to either vary by only a few days within a given month (e.g., the Thanksgiving, the Labor and Memorial Days) or be fixed (e.g., the Christmas). So their effects on the CPI are well absorbed in a calendar month and are covered in the regular seasonality. One exception is the Easter. However, Bureau of Labor Statistics in the US has not considered any holiday effect in its seasonal adjustment procedure for the CPI-US (BLS, 2007), which indicates that the Easter Holiday may have little impact on the CPI-US. Our analysis in Section 5.1 also confirms that CPI-US has no significant holiday effects.

# 3 Linear Time Series Modeling

The analysis of McCracken and Ng (2016) has shown that the auto-regressive (AR) models were adequate for CPI-US series in terms of the in-sample fit and the out-of-sample forecasts. They found that the conventional times series models were not easily out-performed by the contemporary state-of-the art models, such as the factor models with diffusion indexes. We therefore first consider the linear time series models for CPI-CN and then compare their performance with the diffusion indexes models.

## 3.1 Seasonal ARIMA (S-ARIMA) Models

Let $\{Z_t\}_{t=1}^T$ denote the CPI-CN series and $B$ be the back-shift operator such that $BZ_t = Z_{t-1}$. As CPI-CN is not seasonally adjusted, and our preliminary analysis in Figure 2 (b) indicates a strong annual cycle, we consider the S-ARIMA $(p, d, q) \times (P, D, Q)_{12}$ models (Box et al., 2016, chap. 9):

$$\phi_p(B)\Phi_P(B^{12})(1-B)^d(1-B^{12})^D Z_t = \theta_q(B)\Theta_Q(B^{12})a_t, \qquad (1)$$

where $\phi_p(B)$ and $\theta_q(B)$ are polynomials in $B$ of orders $p$ and $q$, respectively, representing the autoregressive (AR) and the moving average (MA) components, $\Phi_P(B^{12})$ and $\Theta_Q(B^{12})$ are polynomials in $B^{12}$ of orders $P$ and $Q$, respectively, representing the seasonal AR and MA components. Roots of the four polynomials are required to lie outside the unit circle to ensure stationarity and invertibility. The $d$ and $D$ are respectively the orders of regular and the seasonal differences. The innovation $\{a_t\}$ is a white noise sequence with mean 0 and finite variance $\sigma_a^2 > 0$.

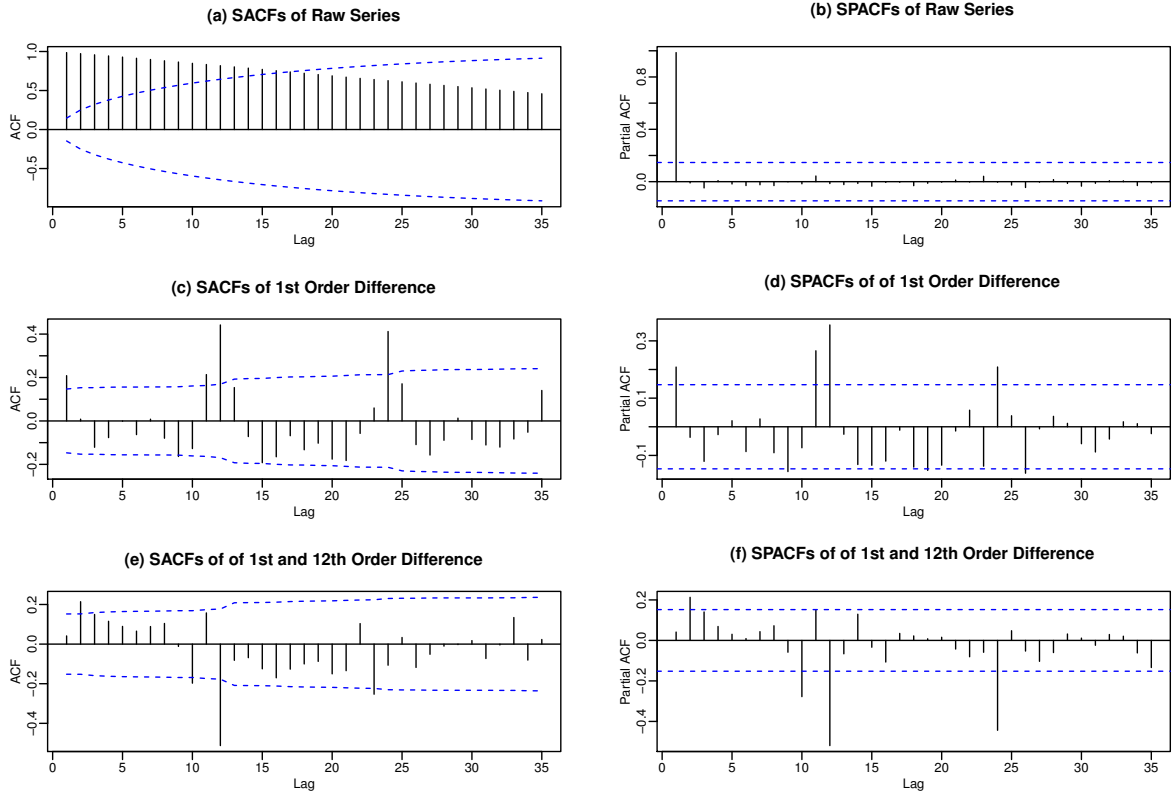

Figure 3: Sample autocorrelation functions (SACF) and sample partial autocorrelation functions (SPACF) for CPI-CN series. Panel (a) and (b): SACF and SPACF of the raw series; Panels (c) an (d): SACF and SPACF of the first differenced series; Panels (e) an (f): SACF and SPACF of the first and the twelfth differenced series.

To determine the orders $p$, $P$, $q$, $Q$, $d$ and $D$, we consider the sample autocorrelation functions (SACFs) and the sample partial autocorrelation functions (SPACFs). Figure 3 displays the SACFs and the SPACFs for CPI-CN, together with those of the first order difference series, and a further twelfth order difference series, respectively. We first try to narrow down the ranges of $d$ and $D$. As the sample ACFs in Figure 3(e) are only significantly none-zero at lags 2, 10 and 12, which suggested taking the first and the twelfth differences are sufficient to make the series stationary. We next carry out the Augmented Dickey-Fuller (ADF) test (Said and Dickey, 1984) for

$H_0 : (1 - B)(1 - B^{12})Z_t$ has unit root versus $H_1$: $(1 - B)(1 - B^{12})Z_t$ is stationary.

The P-value of the test is 0.01, which suggests the unit root hypothesis should be rejected and there is quite amount of evidence to support the stationarity of the CPI series after taking the first and the twelfth differences. This together with the evidence from the SACF and SPACF plots in Figure 3 suggests that $d = D = 1$ in (1), which leads to the following S-ARIMA model

$$\phi_p(B)\Phi_P(B^{12})(1 - B)(1 - B^{12})Z_t = \theta_q(B)\Theta_Q(B^{12})a_t. \qquad (2)$$

It remains to select the ranges for $p$, $q$, $P$ and $Q$ in Model (2), which then allows us to carry out an exhaustive search for the top models in Sub-section 3.4. The SACFs of $(1 - B)(1 - B^{12})Z_t$ shown in Figure 3(e) are only significant at lags 2, 10, 12 and 23, which indicates the MA order $q$ and seasonal MA order $Q$ may be 1 or 2. The SPACFs in Figure 3(f) are only significant at lags 2, 10, 12 and 24, which indicates the AR order $p$ and seasonal AR order $P$ may also be 1 or 2. Hence, we choose $p, q, P$ and $Q$ from $\{0, 1, 2\}$ respectively, resulting in a total of 81 candidate models for (2).

## 3.2 Outlier Detection

Before we proceed with finding out the top performing S-ARIMA models for CPI-CN, we try to detect outliers in the series. The outlier detection is a way to check on the quality of a price series to see if it can react naturally to unexpected events.

We employ the procedure of Chen and Liu (1993) to each of the 81 candidate models for potential outliers. The procedure can detect four types of outliers: Additive Outlier (AO), Innovation Outlier (IO), Level Shift (LS) and Transient Change (TC), which are naturally implemented with the S-ARIMA model (2). To be precise, let $\Psi(B) = [\theta_q(B)\Theta_Q(B^{12})] / [\phi_p(B)\Phi_P(B^{12})(1 - B)(1 - B^{12})]$, then Model (2) can be written as

$$Z_t = \Psi(B)a_t. \qquad (3)$$

The four types of outliers can be augmented with the S-ARIMA model as follows,

$$\text{Additive Outlier:} \quad Z_t = \Psi(B)a_t + \omega P_t^{(t_1)}, \tag{4}$$

$$\text{Innovation Outlier:} \quad Z_t = \Psi(B)(a_t + \omega P_t^{(t_1)}), \tag{5}$$

$$\text{Level Shift:} \quad Z_t = \Psi(B)a_t + \frac{\omega}{1-B}P_t^{(t_1)}, \tag{6}$$

$$\text{Transient Change:} \quad Z_t = \Psi(B)a_t + \frac{\omega}{1-\delta B}P_t^{(t_1)}, \tag{7}$$

where $P_t^{(t_1)} = I(t = t_1)$ represents the impulse function at $t = t_1$, $I$ is the indicator function and $\omega$ indicates the strength of the outlier effect.

The parameters of model (2) can be estimated by the maximum likelihood method assuming the innovations $a_t \sim N(0, \sigma_a^2)$, which gives the estimate of $\Psi(B)$, denoted as $\hat{\Psi}(B)$. Then, we obtain the estimated residuals $\hat{e}_t := Z_t/\hat{\Psi}(B)$, which are subject to the influence of outliers. Thus, for each type of outliers,

$$\hat{e}_t = \omega x_t + a_t, \tag{8}$$

where for the AO type: $x_t = \hat{\pi}(B)P_t^{(t_1)}$; for IO: $x_t = P_t^{(t_1)}$; for LS: $x_t = \hat{\pi}(B)P_t^{(t_1)}/(1-B)$ and for TC: $x_t = \hat{\pi}(B)P_t^{(t_1)}/(1-\delta B)$ with $\delta = 0.8$ as suggested in Chen and Liu (1993). The outlier detection is performed via the $t$-test on the significance of $\omega$ via the linear regression of $\hat{e}_t$ on $x_t$.

Figure 4 displays the evolution of the outliers from the basic model (2) to the one after taking into account of the Spring Festival (SF) and a wide-spread snow storm in Southern China in 2008. The dramatic reduction in the number of outliers suggests that accounting for the SF and the snow event in CPI-CN is necessary. Figure 4 (a) provides monthly frequencies of models (among the 81 candidate models of (2)) that report outliers for CPI-CN before considering the SF and other events. It shows that 96% of outliers are located in Januaries and Februaries. As the SF occurs either in January or February, this pattern of outliers suggests a strong presence of the SF effect on the CPI-CN, which leads us to consider the SF effect in the following.

---

[6] For instance in Panel (a), the first spike from the left is for February 2004 with frequency 72, which means there were 72 models out of the 81 models detected outliers for the month. The legend to this panel lists the number of outliers reported by the 81 models (474), and the percentage of outliers located on January (32%), February (64%) and other months (4%).

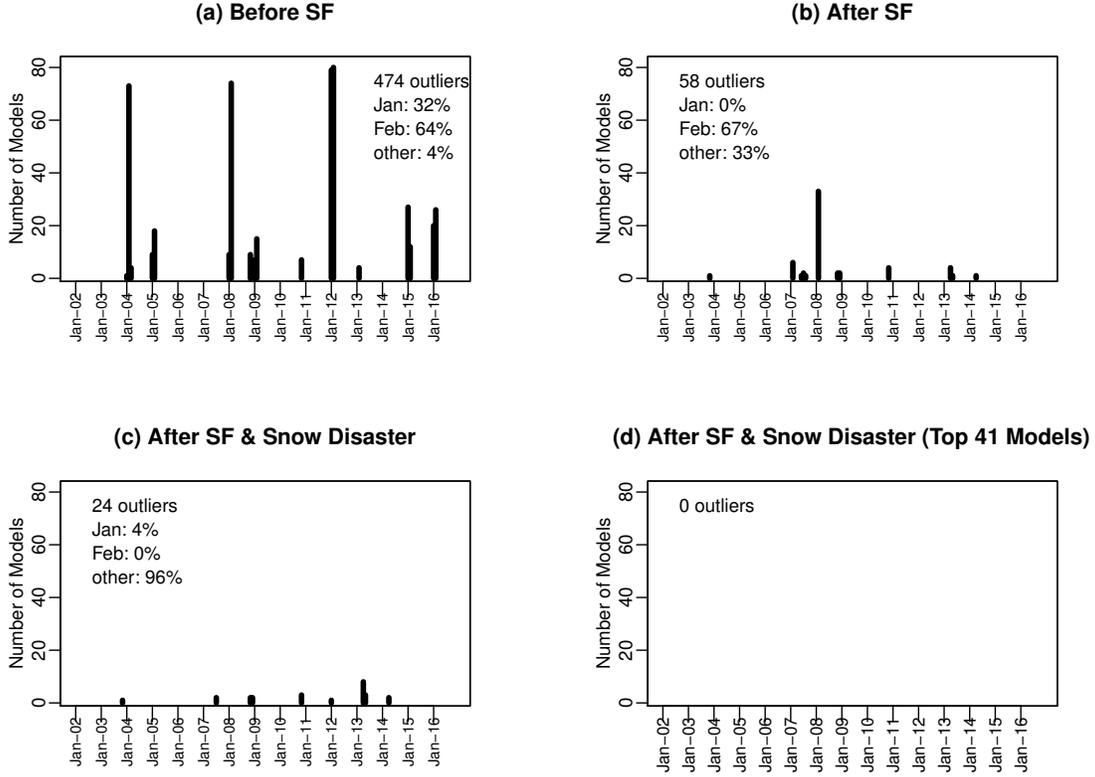

Figure 4: Monthly frequencies of models (out of the 81 models of (2)) that contain outliers for CPI-CN on each month between January 2002 to November 2016. Panel (a): before considering the SF effects; Panel (b): after considering the SF effects; Panel (c): after considering the SF effects and the snow event on January and February 2008; Panel(d): the results of first group (top 41 models selected in Sub-section 3.4) after considering the SF effects and the snow disaster in 2008. The legend on each panel shows the total number of outliers reported, and the percentage of outliers located on January, February and other months, respectively [6].

## 3.3 Spring Festival Effects and Outliers

Consumer prices tend to increase around the SF due to strong demand on consumer goods during the most important holiday in China, while the supply, especially that on food items, is relatively low as the SF is in the winter, a time when the temperature is low for food production and workers are in vocation for family union. The major challenge to study the SF's effect is that the date of the lunar new year differs in the solar calendar every year, ranging from January 21 to February 20. The fluctuation of the date creates different lengths (in terms of number of days) of influence on the price index in January, February and March, which have to be accounted accordingly.

Bell and Hillmer (1983) and Lin and Liu (2003) accounted for the the SF effect on a month by the length of days influenced by the SF. Typically, the economic activity surges before the SF, perks during the SF and slowly dissipate after the SF, while the consumption by consumers have different patterns in the three periods. Hence, we partition the whole SF period into three sub-periods: before period of length $\tau_1$ days, during period of length $\tau_2$ days and after period of length $\tau_3$ days. The three length parameters need to be determined from the data.

Suppose the SF occurs on the $d_0$-th day of the solar calender month. Then, the period between day $d_0 - \tau_1$ and day $d_0 - 1$ is the "before" period, that between day $d_0$ and day $d_0 + \tau_2 - 1$ is the "during" period, and that between day $d_0 + \tau_2$ and day $d_0 + \tau_2 + \tau_3 - 1$ is the "after" period. For a given month $t$, we can calculate the number of days of the month that fall into each of the three periods. Let $\tau_{1t}$, $\tau_{2t}$ and $\tau_{3t}$ be the number of days in month $t$ that intersect with the three periods respectively, and we use

$$H_{it}(\tau_i) = \frac{\tau_{it}}{\tau_i} \quad \text{for} \quad i = 1, 2 \text{ and } 3 \tag{9}$$

to measure the three SF effects[7]. Incorporating them into model (2), we obtain

$$\phi_p(B)\Phi_P(B^{12})(1-B)(1-B^{12})(Z_t - \sum_{i=1}^{3}\beta_i H_{it}(\tau_i)) = \theta_q(B)\Theta_Q(B^{12})a_t. \tag{10}$$

Since $\tau_i$ ($i = 1, 2, 3$) are unknown, we tune them along with the model orders $(p, q, P, Q)$ based on the model fitting and forecasting criteria that will be reported in Sub-section 3.4.

Figure 4(b) shows the counts of models with outliers after taking the SF effects into consideration by model (10). We can see that the total number of outliers is dramatically reduced from 575 to 58, which indicates both the profound SF effects on the CPI-CN and the effectiveness of the approach that accounts for the SP effects. However, there is still a cluster of outliers remained in February 2008. This is a time when a severe snow event, which started from the middle of January and lasted till the end of February 2008, struck a large region in Southern China, where produces a substantial amount of pork, eggs, vegetables and other fresh products (especially in winter) for the domestic market. During the prolonged event, the prices of many items, especially for food items, had increased substantially over entire China and Southern China in particular. Since it was an one-off event, we treat it as an AO type outlier by adding a term $\omega P_t^{(2008-02)}$ to (10), which leads to the following model for CPI-CN:

$$\phi_p(B)\Phi_P(B^{12})(1-B)(1-B^{12})(Z_t - \omega P_t^{(2008-02)} - \sum_{i=1}^{3}\beta_i H_{it}(\tau_i)) = \theta_q(B)\Theta_Q(B^{12})a_t, \quad (11)$$

---

[7]Here, for convenience, we assume that the effects depends linearly on the number of days in each sub-period. See Bell and Hillmer (1983) and Lin and Liu (2003) for a similar treatment.

After considering the snow event, there were only 24 outliers left for the 81 models over the 15 year period, and no outliers for the top 41 performing models as shown in Figure 4 (c) and (d)) respectively. Hence, we regard Model (11) as the final linear time series model, and is referred to as S-ARIMAX.

## 3.4 Model selection

There are two aspects in model selection within the family of S-ARIMAX model (11). One is the selection of $(p, q, P, Q)$ from the 81 possible choices; and the other is selecting the SF effects $\tau_i$, $i = 1, 2, 3$. In order to relieve the computation, we restrict the lengths of the three SF periods such that $\tau_1, \tau_3 \in \{0, 4, 8, 12, 16, 20, 24\}$ and $\tau_2 \in \{0, 4, 8\}$, which leads to $3 \times 7^2 = 147$ combinations. In total, there are $81 \times 147$ possible model choices.

The selection of these parameters is based on the model performance with different choices of $(p, q, P, Q)$ and $(\tau_1, \tau_2, \tau_3)$. Let $\boldsymbol{\tau} = (\tau_1, \tau_2, \tau_3)'$ and $\hat{\sigma}_a(\boldsymbol{\tau}; p, q, P, Q)$ be the maximum likelihood estimator of $\sigma_a$ in Model (11) by assuming that the innovations $\{a_t\}$ are IID $N(0, \sigma_a^2)$. We use

$$C_{Fit}(\boldsymbol{\tau}; p, q, P, Q) = \hat{\sigma}_a(\boldsymbol{\tau}; p, q, P, Q)/\hat{\sigma}_0 \quad (12)$$

to measure the fitting performance of the model. Here $\hat{\sigma}_0$ is the standard error of series $(1-B)(1-B^{12})CPI_t$ as a measure of the baseline variation for each series. For CPI-CN, the baseline variation is $\hat{\sigma}_0 = 0.7244$, and that for CPI-US is $\hat{\sigma}_0 = 1.0348$. We also consider

Table 1: Frequencies of SF effects for top 41 models for CPI-CN.

| SF Effects $(\tau_1, \tau_2, \tau_3)$ | (4, 8, 12) | (4, 0, 12) | (4, 0, 16) | (8, 8, 12) |
|---|---|---|---|---|
| Top group (41 Models) | 18 | 11 | 5 | 5 |
| SF Effects $(\tau_1, \tau_2, \tau_3)$ | (8, 0, 16) | (4, 8, 0) | | |
| Top group (41 Models) | 1 | 1 | | |

the Bayesian Information Criterion (BIC) (Schwarz, 1978) that balances both model fit and model complexity

$$BIC(\boldsymbol{\tau}; p, q, P, Q) = -2\log(L(\boldsymbol{\tau}; p, q, P, Q)) + \log(n)k, \tag{13}$$

where $L(\boldsymbol{\tau}; p, q, P, Q)$ is the maximum likelihood of Model (11) based on the $(1-B^{12})(1-B)CPI_t$ which has $n = T - 13$ observations, $k = p + q + P + Q + |\boldsymbol{\tau}|$ is the total number of parameters excluding $\sigma_a$, and $|\boldsymbol{\tau}|$ is the number of non-zero elements in $\boldsymbol{\tau}$.

Define $RMSE(\boldsymbol{\tau}; p, q, P, Q) = \{\sum_{t=85}^{179}(C\hat{P}I_t - CPI_t)^2/83\}^{1/2}$ to be the 1-step (month) expanding window forecasting Root of Mean Squared Error (RMSE) with the forecasting horizon from January 2009 to November 2016 (95 months) while the training data are from January 2002 to December 2008 (84 months). We use

$$C_{FC}(\boldsymbol{\tau}; p, q, P, Q) = RMSE(\boldsymbol{\tau}; p, q, P, Q)/\hat{\sigma}_0 \tag{14}$$

as the criterion for the forecasting performance.

The model selection procedure is conducted in two steps. In the first step, for each of the 81 $(p, q, P, Q)$-choices in S-ARIMAX Models (11), we obtain $C_{Fit}(\boldsymbol{\tau}; p, q, P, Q)$, $BIC(\boldsymbol{\tau}; p, q, P, Q)$ and $C_{FC}(\boldsymbol{\tau}; p, q, P, Q)$ for all 147 SF choices of $\boldsymbol{\tau} = (\tau_1, \tau_2, \tau_3)'$, and acquire their respective ranks in ascending order among the 147 choices. We choose the $\boldsymbol{\tau}$ that produces the smallest sum of ranks under the three criteria, denoted as $\boldsymbol{\tau}^*(p, q, P, Q)$, which is the optimal SF choice for the given $(p, q, P, Q)$. Then each model has three criteria with respect to its optimal SF choice: $C_{Fit}(p, q, P, Q) := C_{Fit}(\boldsymbol{\tau}^*; p, q, P, Q)$, $BIC(p, q, P, Q) := BIC(\boldsymbol{\tau}^*; p, q, P, Q)$ and $C_{FC}(p, q, P, Q) := C_{FC}(\boldsymbol{\tau}^*; p, q, P, Q)$. In the second step, we choose the optimal $(p, q, P, Q)$ that has the smallest sum of ranks under these three criteria.

Figure 5 shows the pair-wise scatter plots of the three model selection criteria of the 81 candidates of S-ARIMAX Models (11), after considering the SF and the 2008 snow event.

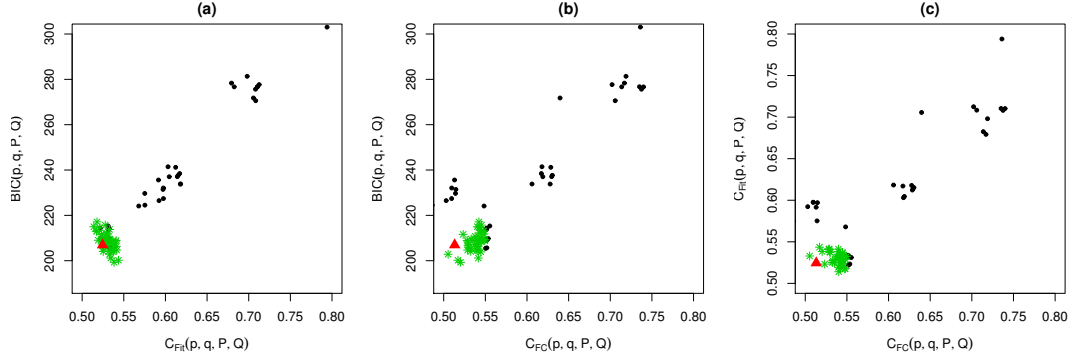

Figure 5: Pairwise scatter plots of the three model selection criteria: $C_{Fit}(p, q, P, Q)$, $BIC(p, q, P, Q)$ and $C_{FC}(p, q, P, Q)$ for CPI-CN based on the 81 candidate models after considering the 147 SF choices and the snow disaster in 2008. The red triangle indicates those of the best model with the smallest sum of ranks, and the green stars indicate the top 41 models.

It is clear that there is a cluster of models that consists of 41 models (the top 41 models) that have outstanding performance under the three criteria so that their $C_{Fit}(p, q, P, Q) \leq 0.55$, $BIC(p, q, P, Q) \leq 220$ and $C_{FC}(p, q, P, Q) \leq 0.55$. Among the top 41 models, the frequencies for the SF effects are listed in Table 1 which shows that most of the top models selected $(\tau_1, \tau_2, \tau_3) = (4, 8, 12)$ or $(\tau_1, \tau_2, \tau_3) = (4, 0, 12)$ as their optimal SF effects.

Table 2 summarizes the numerical values of the three criteria for the top eight models and their respective ranks. It is clear that no model is uniformly better than others under the three criteria. Hence, we use the sum of the three ranks as the overall criterion for model selection. Specifically, S-ARIMAX$(1, 1, 0) \times (1, 1, 2)_{12}$

$$(1 - \phi B)(1 - \Phi B^{12})(1 - B)(1 - B^{12})(Z_t - \omega P_t^{(2008-02)} - \sum_{i=1}^{3} \beta_i H_{it}(\tau_i)) = (1 - \Theta_1 B^{12} - \Theta_2 B^{24}) a_t, \qquad (15)$$

with $(\tau_1, \tau_2, \tau_3) = (4, 0, 12)$ stands out as the best model. The model reports no outlier for the entire data range, with the overall SF effects $\sum_{i=1}^{3} \hat{\beta}_i H_{it}(\tau_i)$ being displayed in Figure 6 (a). To show more clearly the significance of the SF effects, Figure 6 (b) plots the relative percentages of the SF effects, which accounts for the SF effect on CPI annual rate:

$$\text{relative percentage of SF effects} = \left| \frac{R_t - \tilde{R}_t}{R_t} \right|, \qquad (16)$$

Table 2: Model fitting and 1-step out-of-sample forecasting results of the top eight candidates for CPI-CN with the corresponding ranks in the parentheses.

| $(p,q,P,Q)$ | $(\tau_1,\tau_2,\tau_3)$ | $C_{Fit}$ | $BIC$ | $C_{FC}$ | Sum of Ranks |
|---|---|---|---|---|---|
| $(1,0,1,2)$ | $(4,0,12)$ | 0.5245 (13) | 206.95 (17) | 0.513 (7) | 37 |
| $(0,1,2,1)$ | $(4,8,12)$ | 0.5253 (16) | 204.11 (6) | 0.533 (18) | 40 |
| $(1,0,2,2)$ | $(4,0,16)$ | 0.5254 (17) | 205.27 (11) | 0.533 (20) | 48 |
| $(0,1,1,2)$ | $(4,8,12)$ | 0.5193 (5) | 209.02 (28) | 0.540 (24) | 57 |
| $(1,0,2,0)$ | $(4,0,16)$ | 0.5383 (46) | 199.28 (1) | 0.520 (11) | 58 |
| $(2,0,1,2)$ | $(4,0,12)$ | 0.5232 (9) | 211.61 (38) | 0.523 (12) | 59 |
| $(1,0,0,1)$ | $(4,0,12)$ | 0.5436 (51) | 200.20 (2) | 0.517 (10) | 63 |
| $(0,2,2,1)$ | $(4,0,12)$ | 0.5307 (24) | 207.10 (18) | 0.537 (22) | 64 |

where $R_t$ is the annual rate of raw CPI-CN at month $t$ and $\tilde{R}_t$ is that of CPI-CN after subtracting the SF effects. The relative percentages of SF effects in January and February range from 3% to 191%, and most of them are between 25% to 75%, which means substantial SF effects among the CPI-CN annual rates. Hence, it is important to adjust the SF effect for the two months.

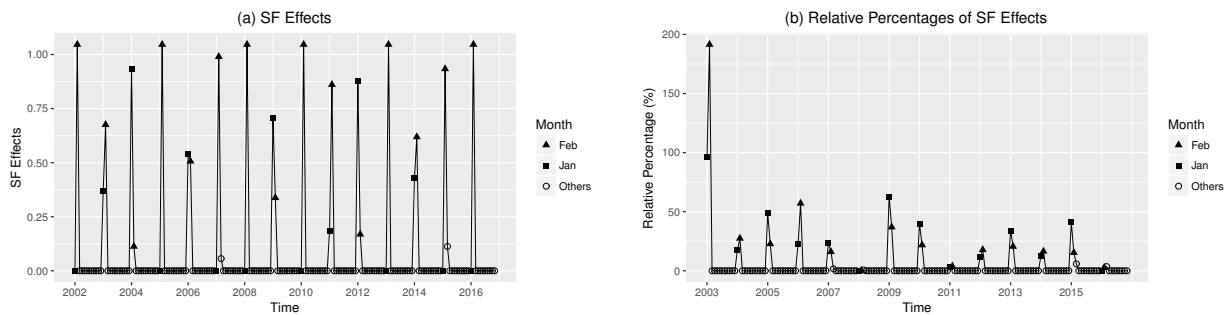

Figure 6: Left panel (a) gives the estimated SF effects $\sum_{i=1}^{3}\hat{\beta}_i H_{it}(\tau_i)$. Right panel (b) shows the relative percentage between CPI annual rate before subtracting SF effects and that after subtracting SF effects.

Table 3 reports the fitting and forecasting performance of the best models for CPI-CN and the CPI-US, respectively. It shows that considering the SF effects and the 2008 snow event for CPI-CN can improve both the fitting and forecasting performance of Model (15).

Table 3: Fitting (Jan. 2002 to Dec. 2008) and forecasting (Jan. 2009 to Nov. 2016) performance of the best S-ARIMAX models for CPI-CN and CPI-US. For CPI-CN, "Before" and "After" means before and after considering SF effects and 2008 snow event, respectively. For CPI-US, "Before" and "After" means before and after accounting for the outliers in Sub-section 5.1, respectively.

|  |  | Fitting | Out-of-Sample Forecasting RMSE/$\hat{\sigma}_0$ |  |  |  |  |  |
|---|---|---|---|---|---|---|---|---|
|  |  | $\hat{\sigma}_a/\hat{\sigma}_0$ | 1-step | 2-step | 3-step | 6-step | 9-step | 12-step |
| CPI-CN | Before | 0.677 | 0.680 | 0.874 | 1.146 | 2.026 | 2.936 | 3.638 |
|  | After | 0.525 | 0.513 | 0.820 | 1.073 | 1.915 | 2.781 | 3.533 |
| CPI-US | Before | 0.576 | 0.572 | 0.882 | 1.288 | 2.425 | 3.152 | 3.410 |
|  | After | 0.412 | 0.491 | 0.866 | 1.267 | 2.346 | 3.023 | 3.274 |

The standard deviation of the fitting errors ($\hat{\sigma}_a/\hat{\sigma}_0$) was reduced by 22.5%, and the one-step out-of-sample forecasting RMSE was decreased by 24.5%. Table A.1 in the Supplementary Material reports 1, 2, 3, 6, 9, 12-step out-of-sample forecast results of Model (15) for CPI-CN. Table 3 also shows by considering outliers the performance of the best model for CPI-US can also be improved. See Sub-section 5.1 for details.

# 4 The Diffusion Index Approach

Our analysis so far has been confined to the univariate modeling of the CPI using the S-ARIMAX models. This univariate modeling approach ignores information contained in other macroeconomic variables that might have predictive content for the CPI. To enhance the forecasting performance, we collect major economic variables and consider utilizing the information contained in these variables as the Diffusion Index (DI) via the principal component analysis (Stock and Watson, 2002a,b; McCracken and Ng, 2016) . We then predict the seasonal adjusted CPI with the estimated DIs, where the number of DIs is selected by the BIC (Schwarz, 1978).

The premise of the DI approach is that the information in the predictors can be represented by a small number of estimated factors. The approach avoids the task of variable selection which may miss the variables that have predictive power in out-of-sample but lack significance for the in-sample fit. This idea has a long tradition in macroeconomics as articulated in Stock and Watson (2002a,b) and references therein. For CPI-CN, we consider 31 main economic variables which are listed in Table A.2 in the Supplementary Material. And for the CPI-US, we use FRED monthly database, which is a proper substitute for Stock-Watson dataset based on the study of McCracken and Ng (2016).

## 4.1 Seasonal Adjustment and Data Pre-processing

The covariates used to build the DI are monthly observations, which have the same time span as that of the CPI-CN. As the officially released Chinese data are not seasonally adjusted, we perform the seasonal adjustment first on each of the covariate series. For a given series that contains the SF effects, say $Z_t$ (sometimes after a logarithm transformation), applying the S-ARIMAX model (11) to the series, we first estimate the SF effects, which can be expressed as $\hat{H}_t = \sum_{i=1}^{3} \hat{\beta}_i H_{it}(\tau_i)$. Subtract this estimated SF effects from $Z_t$, we obtain the SF-removed series $\tilde{Z}_t = Z_t - \hat{H}_t$. The X-13ARIMA-SEATS procedure (U.S. Census Bureau, 2017) is applied to this $\tilde{Z}_t$ to decompose it into trend $T_t$, seasonality $S_t$ and irregular component $I_t$, and finally obtain the seasonal adjusted series $saZ_t$, using either additive model or multiplicative model as below

$$\begin{aligned} \text{Additive model:} \quad & \tilde{Z}_t = T_t + S_t + I_t, \quad saZ_t = \tilde{Z}_t - S_t = Z_t - \hat{H}_t - S_t, \\ \text{Multiplicative model:} \quad & \tilde{Z}_t = T_t S_t I_t, \quad saZ_t = \tilde{Z}_t / S_t = (Z_t - \hat{H}_t)/S_t, \end{aligned} \quad (17)$$

depending on the nature of the covariate.

After the seasonal adjustment, we apply proper transformations on these seasonal adjusted series to make them stationary. Table A.2 in Supplementary Material provides details on the pre-treatments applied to each covariate. Finally, we standardize these transformed series to form multivariate time series $\boldsymbol{X}_t$ for the factor analysis needed in the DI analysis. In particular, we denote the seasonally adjusted CPI-CN by $saCPI_t$.

## 4.2 Diffusion Indexes Estimation

After the data pre-processing, we built the following model to estimate the Diffusion Indexes (DIs) from $n \times 1$ stationary multivariate time series $\boldsymbol{X}_t$:

$$\boldsymbol{X}_t = \boldsymbol{\Lambda} \boldsymbol{F}_t + \boldsymbol{e}_t, \tag{18}$$

where $\boldsymbol{F}_t$ is the $r \times 1$ common factors, $\boldsymbol{\Lambda}$ is the $n \times r$ matrix of factor loadings, and $\boldsymbol{e}_t$ is the vector of idiosyncratic components satisfying $E(\boldsymbol{e}_t|\boldsymbol{F}_t) = \boldsymbol{0}$ and finite second moments. Here the common factors $\boldsymbol{F}_t$ are estimated by the Principle Component Analysis (Anderson, 2003). According to Stock and Watson (2002b), the estimated factors can be interpreted in terms of diffusion indexes developed by NBER business cycle analysts to measure common movement in a set of macroeconomic variables. Stock and Watson (2002a) established that the principal component estimator is point-wise (for any date $t$) consistent and has limiting mean squared error (MSE) over all $t$ that converges to 0, under a suitable set of regularity conditions.

Some series in our dataset, like the annual rate of Chinese production price index (PPI) for the light manufacture industry, contain missing observations. As a result, the standard principal components analysis does not apply. However, the expectation maximization (EM) algorithm of Dempster et al. (1977) can be applied by solving a minimization problem iteratively; see the Appendix of Stock and Watson (2002b) for details of the algorithm to estimate $\boldsymbol{F}_t$.

## 4.3 Forecasting with the DIs

The multi-period predictions are projections of an $h$-step-ahead variable $y_{t+h}^h$ onto $t$-dated predictors and lagged values of $y_t$. The logarithm of $saCPI$ is I(2) non-stationary, we followed the procedure in Stock and Watson (2002b) and define

$$y_{t+h}^h := (1200/h)\ln(saCPI_{t+h}/saCPI_t) - 1200\ln(saCPI_t/saCPI_{t-1}) \tag{19}$$

and $y_t := 1200 \triangle \ln(saCPI_t/saCPI_{t-1})$.

The general forecasting equation for the dynamic DI approach is

$$y_{t+h}^h = \alpha + \sum_{j=0}^{\tilde{p}-1} \gamma_j y_{t-j} + \sum_{j=0}^{m-1} \tilde{\boldsymbol{\beta}}_j' \boldsymbol{f}_{t-j} + \varepsilon_{t+h}, \tag{20}$$

Here $\tilde{p}$ and $m$ are orders of the auto-regressive lags and lagged factors respectively, and $\boldsymbol{f}_t$ which is $k \times 1$ vector for $k \leq r$ is a subset of $\boldsymbol{F}_t$ estimated from the Factor Model (18). To forecast $h$-step ahead of $y_{t_0+h}^h$ at a time $t_0$ (training dataset ends at $t_0$), we have three tunning parameters: number of factors $k$, lags of $y_t$, $\tilde{p}$, and lags of the factors $m$. Denote $max.k$ as the largest possible number of factors used in forecasting. For each $max.k$ in $\{1,2,3,4,5\}$, we choose $(k, \tilde{p}, m) \in \{1 \leq k \leq max.k, 1 \leq \tilde{p} \leq 6, 1 \leq m \leq 6\}$ based on the BIC; then use the selected $k$, $\tilde{p}$ and $m$ to fit model (20) and carry out the forecasting of $y_{t_0+h}^h$, denoted as $\hat{y}_{t_0+h}^h$. For CPI-CN, $\tilde{p}$ was mostly selected as 4, 5 and 6, $m$ mostly 1,2,3, $k$ mostly 1 and 2. For CPI-US, $\tilde{p}$ was always selected as 6, $m$ mostly 1 and sometimes 3, $k$ mostly 1 and 2.

After obtaining $\hat{y}_{t_0+h}^h$, via formula (19), we can get the $h$-step ahead forecast of the seasonal adjusted CPI, $sa\hat{CPI}_{t_0+h}^h = \exp(h\hat{y}_{t_0+h}/1200)\, saCPI_{t_0}^{h+1}/saCPI_{t_0-1}^h$. Then we predict seasonality $S_{t_0+h}$ via S-ARIMA whose orders are selected based on SACFs and SPACFs of $\{S_t\}$. Finally the prediction of $CPI_{t_0+h}^h$ is obtained from $\hat{CPI}_{t_0+h}^h = sa\hat{CPI}_{t_0+h}^h + \hat{S}_{t_0+h} + \hat{H}_{t_0+h}$.

To compare the forecasting performance of the DI approach with that of the S-ARIMAX model, we carry out the out-of-sample forecast for the period between January 2009 and November 2016, with the forecasting horizons $h = 1, 2, 3, 6, 9$ and 12 months respectively. Here we consider two forecasting schemes, i.e., the expanding window and the rolling window respectively. In the expanding window scheme with forecast horizon $h$, the training dataset is the monthly CPI-CN series from January 2002 till $h$-month prior to the month $t$ where the forecast is desired. In the rolling window scheme, the training data-set is the moving window of $(49 - h)$ months prior to the month $t$.

## 4.4 DI Result for CPI-CN

We apply the procedure in the above two sub-sections to forecast seasonal adjusted CPI-CN, and use S-ARIMA $(0, 0, 1) \times (0, 1, 1)_{12}$ for modeling the seasonality of $S_t$. Table 4 collects the ratio of out-of-sample forecasting RMSE of the DI approach over that of the S-ARIMAX. It is observed that the DI approach significantly improves upon the S-ARIMAX model in the 1, 2 and 3-step forecasts for the expanding window method, reducing the forecasting RMSE by 10%, 14% and 9%, respectively. For the rolling window method, the DI only improves upon the S-ARIMAX model in the 1 and 2-step forecasts. This may be due to the rolling window method employed smaller number of observations than the expanding window, which makes its DI estimation less stable than that of the expanding window. The improvements are also reflected in Figure A.1 (a)-(f) in the Supplementary Material which shows that the forecasting errors of the DI approach are smaller than those of the S-ARIMAX model after year 2010.

Table 4: Ratio of out-of-sample forecasting RMSE of DI approach (18 & 20) over that of S-ARIMAX model (15) for the CPI-CN from January 2009 to November 2016. Value smaller than 1 means DI approach has smaller forecasting RMSE than S-ARIMAX. Forecasting horizon $h$ ranges from 1 (one month) to 12 (one year), and $max.k$ is the largest possible number of factors used in forecasting.

| | Expanding Window | | | | | |
|---|---|---|---|---|---|---|
| Forecast Horizon | h=1 | h=2 | h=3 | h=6 | h=9 | h=12 |
| $max.k = 1$ | 0.90 | 0.87 | 0.94 | 1.16 | 1.39 | 1.52 |
| $max.k = 2$ | 0.90 | 0.86 | 0.91 | 1.00 | 1.27 | 1.38 |
| $max.k = 3$ | 0.90 | 0.86 | 0.91 | 1.00 | 1.27 | 1.38 |
| $max.k = 4$ | 0.90 | 0.86 | 0.91 | 0.98 | 1.30 | 1.37 |
| $max.k = 5$ | 0.90 | 0.86 | 0.91 | 0.98 | 1.30 | 1.37 |
| | Rolling Window | | | | | |
| Forecast Horizon | h=1 | h=2 | h=3 | h=6 | h=9 | h=12 |
| $max.k = 1$ | 0.89 | 0.92 | 1.03 | 1.31 | 1.49 | 1.59 |
| $max.k = 2$ | 0.89 | 0.94 | 1.00 | 1.30 | 1.47 | 1.51 |
| $max.k = 3$ | 0.89 | 0.94 | 1.00 | 1.32 | 1.47 | 1.50 |
| $max.k = 4$ | 0.89 | 0.94 | 1.00 | 1.31 | 1.49 | 1.49 |
| $max.k = 5$ | 0.89 | 0.93 | 1.00 | 1.31 | 1.49 | 1.49 |

For the 1, 2 and 3-step forecasts, using only the first factor appeared to be sufficient, since adding the second to the fifth factors does not offer obvious improvement for the prediction. This suggests that the first factor may be used as the short-term leading index of CPI-CN. Table A.3 in the Supplementary Material shows that the first factor was mainly composed of the following covariates (id, loading) if we focus on those with the loading bigger than 0.2: PPI (ppi, 0.33), PPI:Living Material (ppi_living, 0.37), PPI:Living Material:Food (ppi_living_food, 0.37); Retail Price Index (rpi, 0.33); Annual Rate of PPI:Light Industry (ppi_light_rate, 0.26); Cash Reserve Ratio (reserve_ratio, 0.26); Rediscount Rate (rediscount_rate, 0.26); Loan Rate (loan_rate, 0.25); CPI:Food:Meat (cpi_food_meat, 0.24); CPI:Food:Grain (cpi_food_grain, 0.22), CPI:Food:Pork (cpi_food_pork, 0.22). Thus, these covariates would have short-term prediction power of the CPI-CN. Moreover, among these top eleven covariates, food occupies four slots, which reflects that the CPI-CN is dominated by food items.

Finally, both Table 4 and Figure A.1 (g)-(l) show that the DI approach has larger forecast errors than the S-ARIMAX model for forecasting horizon $h = 6$, 9 and 12 months. This shows that the DI has less prediction power for the long-term forecast for the CPI-CN, when the S-ARIMAX is more advantageous.

# 5   Results for US CPI Series

In the previous two sections, we have applied the linear time series models and the diffusion index model to study the Chinese CPI series. In the literature, such analysis has been performed for the US CPI series (CPI-US), in earlier studies such as Stock and Watson (2002a,b), and more recent studies like McCracken and Ng (2016). A natural question of central importance is whether the dynamic structure of the CPI-CN and that of the CPI-US would be similar. Furthermore, would both series be predictable? To answer these questions, this section will perform the analysis for the CPI-US parallel to that for the CPI-CN in Section 3 and 4. Our findings for the CPI-US will be briefly reported, for the sake of space, as the analysis is quite similar to those presented in the previous sections. The linkage of our results with those reported in the literature will be highlighted. A discussion in Section 6 summarizes the major findings in the comparison between CPI-CN and CPI-US.

## 5.1 Linear Time Series Modeling

For the CPI-US, Figure A.2 in Supplementary Material provides the SACFs and SPACFs of the series and those after taking the first order and the 12-th order differences, namely for $(1-B)Z_t$ and $(1-B)(1-B^{12})Z_t$ if we denote the raw CPI-US as $Z_t$. Similar to the analysis of CPI-CN, Figure A.2 suggests the following model class for CPI-US, namely

$$\phi_p(B)\Phi_P(B^{12})(1-B)(1-B^{12})Z_t = \theta_q(B)\Theta_Q(B^{12})a_t. \tag{21}$$

which is the same as the model class (2) for CPI-CN.

To select the orders $p$, $q$, $P$ and $Q$ in Model (21), we choose a range based on the SACFs and SPACFs of $(1-B)(1-B^{12})Z_t$ as conveyed in Figure A.2. As the SACFs in Figure A.2(c) are only significant at lags 1, 12, 13, and SPACFs in Figure A.2(f) are significant at lags 1, 2, 12, 15, 24 and 25, we choose these orders $p$, $q$, $P$ and $Q$ from the same range $\{0, 1, 2\}$, making a total of the same 81 candidate models (2) for the CPI-CN.

Figure A.3 in Supplementary Material shows the outlier detection results for the 81 candidate models. As displayed in the Panel (a), there were seven outliers detected by most of the 81 models, three IO type on September 2005, October and November 2008, respectively; two LS on November 2007 and January 2015, and two AO on June 2009 and one TC on February 2013. The IO on September 2005 can be attributed by Hurricane Rita occurred at the end of the August and Katrina in the September of 2005. The LS on January 2015 was caused by a sharp decrease of gasoline price, which reflected the CPI-US is highly influenced by the energy prices. The IO on October and November 2008 can be attributed to the Financial Crisis. Finally, the AO on June 2009 was likely due to the Car Allowance Rebate System (C.A.R.S), which is stated in BLS official intervention analysis (BLS, 2017):"...Car Allowance Rebate System initiative led to a lower supply of used cars and an increase in used car prices...".

Incorporating these seven significant outliers, Model (21) is revised as

$$\phi_p(B)\Phi_P(B^{12})(1-B)(1-B^{12})\tilde{Z}_t =$$
$$\theta_q(B)\Theta_Q(B^{12})(a_t + \omega_1 P_t^{(2005-09)} + \omega_2 P_t^{(2008-10)} + \omega_3 P_t^{(2008-11)}) \tag{22}$$

where $\tilde{Z}_t = Z_t - \omega_1 S_t^{(2007-11)} - \omega_2 P_t^{(2009-06)} - \frac{\omega_3}{1-\delta B} P_t^{(2013-02)} - \omega_4 S_t^{(2015-01)}$. The outlier detection result under this new Model (22) is shown in Figure A.3 (b), which indicates that there are no more outliers in more than half of the candidate models, especially among the top performing models as shown in the following.

Table 5: Model fitting and 1-step out-of-sample forecasting results of the top eight candidates for CPI-US with the corresponding ranks in the parentheses.

| $(p, q, P, Q)$ | $C_{Fit}$ | BIC | $C_{FC}$ | Sum of Ranks |
|---|---|---|---|---|
| $(2, 2, 0, 1)$ | 0.412 (4) | 287.56 (3) | 0.491 (1) | 8 |
| $(2, 2, 1, 1)$ | 0.408 (2) | 291.38 (10) | 0.500 (2) | 14 |
| $(2, 0, 0, 1)$ | 0.427 (14) | 287.12 (2) | 0.500 (3) | 19 |
| $(2, 0, 1, 1)$ | 0.425 (12) | 291.38 (9) | 0.506 (6) | 27 |
| $(2, 0, 0, 2)$ | 0.424 (11) | 291.22 (8) | 0.511 (11) | 30 |
| $(2, 2, 1, 2)$ | 0.410 (3) | 297.55 (22) | 0.508 (9) | 34 |
| $(0, 1, 1, 1)$ | 0.430 (26) | 290.41 (6) | 0.506 (5) | 37 |
| $(2, 2, 0, 2)$ | 0.408 (1) | 290.73 (7) | 0.579 (31) | 39 |

To choose the proper orders $p$, $q$, $P$ and $Q$ in Model (22), we calculate the three criteria ($C_{Fit}(p,q,P,Q) = \hat{\sigma}_a/\hat{\sigma}_0$, BIC$(p,q,P,Q)$ and $C_{FC}(p,q,P,Q) = RMSE/\hat{\sigma}_0$) for the 81 candidate models, shown in Figure A.4. It is clear that there is a cluster of 20 models (the top 20 models marked as green stars in Figure A.4) that stands out clearly based on all three criteria: $C_{Fit}(p,q,P,Q) \leq 0.45$; BIC$(p,q,P,Q) \leq 306$; $C_{FC}(p,q,P,Q) \leq 0.55$. Figure A.3 (c) follows up on the outlier detection results for these top 20 models, which shows a substantial reduction in the outlier counts of the models. The ratio of the number of outliers per model dropped from 3.05 among the 81 models to 1.7 among the top 20 models. It was further reduced to 1.375 among the top 8 models as specified in Table 5. As a result, we only consider the top eight models in the following.

The best performing model with order $(2,1,2) \times (0,1,1)_{12}$ is indicated as the red triangle in Figure A.4, which admits the form

$$(1 - \phi_1 B)(1 - \phi_2 B^2)(1 - B)(1 - B^{12})\tilde{Z}_t =$$
$$(1 - \theta_1 B)(1 - \theta_2 B^2)(1 - \Theta_1 B^{12})(a_t + \omega_1 P_t^{(2005-09)} + \omega_2 P_t^{(2008-10)} + \omega_3 P_t^{(2008-11)}). \quad (23)$$

As shown in Table 3, after considering outliers, the fitted error standard deviation ($\hat{\sigma}_a$) and 1-step out-of-sample forecasting RMSE (relative to $\hat{\sigma}_0$) are reduced by 28% and 14% respectively. Figure A.5 in the Supplementary Material shows that, after considering outliers in Model (23), all the large errors caused by outliers are removed.

Comparing the top eight S-ARIMAX models for CPI-CN and CPI-US, we find the CPI-US tends to choose larger auto-regressive order than the CPI-CN as we note that for CPI-CN $(p,q) = (1,0)$ four times, (0,1) twice, (2,0) and (0,2) once each, while for CPI-US $(p,q) = (2,2)$ four times, (2,0) three times, and (0,1) once. For the (P,Q) order, CPI-CN has (1,2) three times, (2,1) twice, and (2,2), (2,0) and (0,1) once, respectively. In contrast, for CPI-US, the (P,Q) orders are (1,1) three times, (0,1) twice, (0,2) twice and (1,2) once. This difference in order choices implies that CPI-CN has longer seasonal dependence, but less inter-yearly dependence than CPI-US does.

## 5.2 DI Approach Results

The United States has its macroeconomic time series seasonally adjusted, including the CPI series. To gain more details on the CPI-US in comparison with the CPI-CN, we consider the raw CPI-US series. We compute its seasonality as $S_t = CPI_t/saCPI_t$, since BLS uses multiplicative model for seasonal adjustment (BLS, 2007). Data transformation following Stock and Watson (2002b) and McCracken and Ng (2016) are similarly adopted. Then we applied the DI approach (see equation (18) and (20)) on $saCPI_t$, and use S-ARIMA $(2,0,0) \times (2,1,0)_{12}$ to model $log(S_t)$. Predictions of these two components were carried out to obtain the the prediction of CPI via $\hat{CPI}^h_{t+h} = sa\hat{CPI}^h_{t+h}\hat{S}_{t+h}$. The DI approach with $max.k = 2$ gives the most precise forecast, and the forecasting results with $max.k = 1, \cdots, 5$ are listed in Table 6.

Table 6 shows that the DI approach offered inferior results to the univariate S-ARIMAX model (23) for CPI-US. This is in a sharp contrast to the results for the CPI-CN, and seems to contradict with the results of Stock and Watson (2002b) that the DI approach was better in forecasting US inflation compared to the auto-regressive (AR) models. One reason may be that the time span January 2002 to November 2016 is too short for factor extraction and DI approach forecasting. The forecast errors shown in Figure A.6 in the Supplementary Material indicate that DI approach oerfrom worse in the year 2009 for $h = 1, 2, 3, 6$, and in years 2009 and 2010 for $h = 9, 12$. This time period coincides with the recent financial crisis.

Table 6: Ratio of out-of-sample forecasting RMSE of DI approach (18 & 20) over that of S-ARIMAX model (23) for the CPI-US from January 2009 to November 2016.

| | Expanding Window | | | | | |
|---|---|---|---|---|---|---|
| h-step Forecast | h=1 | h=2 | h=3 | h=6 | h=9 | h=12 |
| DI ($max.k = 1$) | 1.35 | 1.71 | 1.97 | 2.18 | 2.19 | 2.18 |
| DI ($max.k = 2$) | 1.26 | 1.76 | 1.94 | 2.18 | 2.19 | 2.18 |
| DI ($max.k = 3$) | 1.32 | 1.73 | 1.92 | 2.34 | 2.17 | 2.18 |
| DI ($max.k = 4$) | 1.32 | 1.73 | 1.90 | 2.39 | 2.14 | 2.18 |
| DI ($max.k = 5$) | 1.32 | 1.73 | 2.07 | 2.77 | 2.14 | 2.18 |
| | Rolling Window | | | | | |
| h-step Forecast | h=1 | h=2 | h=3 | h=6 | h=9 | h=12 |
| DI ($max.k = 1$) | 1.19 | 1.47 | 1.75 | 2.11 | 2.24 | 2.47 |
| DI ($max.k = 2$) | 1.15 | 1.55 | 1.74 | 2.10 | 2.24 | 2.46 |
| DI ($max.k = 3$) | 1.18 | 1.52 | 1.77 | 2.31 | 2.28 | 2.62 |
| DI ($max.k = 4$) | 1.18 | 1.57 | 1.76 | 2.51 | 2.46 | 2.65 |
| DI ($max.k = 5$) | 1.18 | 1.57 | 1.84 | 2.73 | 2.46 | 2.68 |

So it is likely that the financial crisis has more severe impact on the DI approach forecasting than S-ARIMAX forecasting. In addition, the plots of the forecast errors suggest that the financial crisis has more severe influence on CPI-US than on CPI-CN.

Moreover, the model's forecasting ability may change over time and vary for different target series. In Stock and Watson (2002b), the forecasting power of the DI and AR model were compared for $y_{t+h}^h$ from January 1970 to December 1998. However, we forecast the raw CPI-US without seasonal adjustment from January 2009 to November 2016. For comparison, we also perform the same forecasting procedure of Stock and Watson (2002b) to predict $y_{t+h}^h$ for the period January 2009 to November 2016. The ratios of RMSE of DI over RMSE of AR for 1,2,3,6,9 and 12-month forecasting are 1.08 1.28 1.33 1.36 1.20 1.06, indicating that the DI approach is inferior to the linear time series model.

In addition, our results seem to be consistent with those reported in McCracken and Ng (2016), where the authors compared the forecasting of $y_{t+h}^h$ from January 2008 to

Table 7: S-ARIMAX Model Forecasting RMSE/$\hat{\sigma}_0$ of CPI-CN and CPI-US. For CPI-CN, results of Model (15) is listed, and for CPI-US, results of Model (23) is listed.

| Forecast Horizon | h=1 | h=2 | h=3 | h=6 | h=9 | h=12 |
|---|---|---|---|---|---|---|
| CPI-CN (expanding window) | 0.51 | 0.82 | 1.07 | 1.92 | 2.78 | 3.53 |
| CPI-CN (rolling window) | 0.52 | 0.78 | 0.99 | 1.78 | 2.59 | 3.32 |
| CPI-US (expanding window) | 0.49 | 0.87 | 1.27 | 2.35 | 3.02 | 3.27 |
| CPI-US (rolling window) | 0.55 | 1.00 | 1.44 | 2.47 | 3.02 | 3.17 |

February 2012. Their results indicated that the DI approach had the same forecasting power as the AR model for the 1-month horizon, but was worse than the AR model for the 12 and 24 month-horizon, respectively. It is worth noting that Stock and Watson (2002b) and McCracken and Ng (2016) only employed the AR model without the MA component. And they used the BIC to select the model order. Here we firstly select potential AR and MA orders based on ACFs and PACFs, and then use three criteria, $C_{Fit}(p,q,P,Q)$ and $BIC(p,q,P,Q)$ for model fitting and $C_{FC}(p,q,P,Q)$ for forecasting, to select the optimal model. This is more aggressive and better for forecasting than just relying on the BIC criterion. In a word, different forecasting period, different target series and different forecasting procedures lead to a conclusion different from that of Stock and Watson (2002b).

# 6  Discussion

Our study represents a systematic analysis on Chinese consumer price index relative to its US counterpart with both the traditional linear time series models and the more modern diffusion index approach. This paper has three major findings. The first one is that the CPI-CN admits regular dynamic structures that can be well modeled by the conventional linear time series models. The fact that the CPI-CN displays pronounced SF effects year after year despite its timing is shifting between January and February indicates the quality of the CPI-CN as it captures a key consumption pattern of the Chinese consumers. Another aspect is the interpretation of the major outliers by natural and economic events.

The second finding is that the CPI-CN and CPI-US can be modeled by a similar linear time series model class, although there are differences in the composition of the top 8 models as the two series possess different dynamic patterns. For the top eight models of CPI-CN in Table 2, regular orders $p$ and $q$ are mostly 1 or 0, and seasonal orders $P$ and $Q$ are mostly 1 and 2. While for the top eight models of CPI-US in Table 5, $p$ and $q$ are mostly 2, and $P$ and $Q$ are mostly 0 and 1. This implies that CPI-CN has shorter regular dependence and longer seasonal dependence than those of the CPI-US, respectively. Indeed, as summarized in Table 7, in term of predictability of CPI-CN and CPI-US, they are comparable based on their forecast RMSE/$\hat{\sigma}_0$. For h=1, CPI-CN is at least as predictable as CPI-US; for h=2,3,6,9, CPI-CN had less prediction error than CPI-US; for h=12, the predictability of CPI-CN is slightly worse than that of CPI-US. This together with the two series having comparable model fitting accuracy (Table 3) leads to a major conclusion: the Chinese series is as predictable as the US series.

The last major finding is that the predictability of the Chinese CPI series can be further improved by utilizing extra macro-economic information through the DIs for shorter terms (one to three months) forecasting. However, for the CPI-US, the DIs appeared not able to gain further prediction accuracy over the S-ARIMAX models.

## Acknowledgements

The authors would like to acknowledge the support from National Bureau of Statistics of China for providing the raw CPI data, and financial support from National Natural Science Foundation of China Grants 71472007, 71532001 and 71671002, and that from Guanghua School of Management and the Center for Statistical Science, Peking University.

# SUPPLEMENTARY MATERIAL

Table A.1: Forecasting RMSE/$\hat{\sigma}_0$ of the top eight models for CPI-CN with forecasting horizon ranging between one month ($h = 1$) and a year ($h = 12$) after considering the SF and the snow event. The smallest entries for each $h$ is marked in bold.

| $(p, q, P, Q)$ | h=1 | h=2 | h=3 | h=6 | h=9 | h=12 |
|---|---|---|---|---|---|---|
| | | | Expanding Window | | | |
| $(1, 0, 1, 2)$ | **0.513** | 0.820 | 1.073 | 1.915 | 2.781 | 3.533 |
| $(0, 1, 2, 1)$ | 0.532 | **0.763** | **0.999** | 1.856 | 2.723 | 3.435 |
| $(1, 0, 2, 2)$ | 0.533 | 0.822 | 1.083 | **1.811** | **2.602** | **3.382** |
| $(0, 1, 1, 2)$ | 0.540 | 0.832 | 1.109 | 2.032 | 2.984 | 3.739 |
| $(1, 0, 2, 0)$ | 0.520 | 0.783 | 1.052 | 1.848 | 2.669 | 3.469 |
| $(2, 0, 1, 2)$ | 0.523 | 0.839 | 1.109 | 1.993 | 2.853 | 3.529 |
| $(1, 0, 0, 1)$ | 0.517 | 0.811 | 1.048 | 1.878 | 2.718 | 3.474 |
| $(0, 2, 2, 1)$ | 0.537 | 0.774 | 1.003 | 1.863 | 2.723 | 3.417 |
| | | | Rolling Window | | | |
| $(p, q, P, Q)$ | h=1 | h=2 | h=3 | h=6 | h=9 | h=12 |
| $(1, 0, 1, 2)$ | 0.518 | 0.780 | 0.986 | 1.779 | 2.593 | 3.318 |
| $(0, 1, 2, 1)$ | 0.515 | **0.748** | 0.965 | 1.794 | 2.708 | 3.423 |
| $(1, 0, 2, 2)$ | 0.686 | 0.808 | 1.054 | 1.875 | 2.749 | 3.517 |
| $(0, 1, 1, 2)$ | 0.552 | 0.773 | 0.995 | 1.879 | 2.727 | 3.463 |
| $(1, 0, 2, 0)$ | 0.520 | 0.789 | 1.078 | 1.930 | 2.840 | 3.646 |
| $(2, 0, 1, 2)$ | 0.514 | 0.775 | 0.999 | 1.806 | **2.579** | **3.226** |
| $(1, 0, 0, 1)$ | **0.500** | 0.772 | 1.019 | 1.859 | 2.687 | 3.448 |
| $(0, 2, 2, 1)$ | 0.534 | 0.751 | **0.955** | **1.755** | 2.600 | 3.308 |

Table A.2: Variables used in DI Approach for CPI-CN and the transformation applied to them.

| Variable | Name | Source | Unit | Seasonally Adjusted | Transformation Applied |
|---|---|---|---|---|---|
| cpi | CPI | NBSC | Index | Y | 1 |
| cpi_food_grain | CPI: Food: Grain | NBSC | Index | N | 1 |
| cpi_food_meat | CPI: Food: Meat | NBSC | Index | Y | 1 |
| cpi_food_pork | CPI: Food: Meat: Pork | NBSC | Index | Y | 1 |
| cpi_food_vege | CPI: Food: Vegetable | NBSC | Index | Y | 2 |
| cpi_food_fruit | CPI: Food: Fruit | NBSC | Index | Y | 1 |
| cpi_tobacco | CPI: Tobacco, Alcohol and related Articles | NBSC | Index | N | 1 |
| rpi | Retail Price Index | NBSC | Index | Y | 1 |
| rpi_food | RPI: Food | NBSC | Index | Y | 1 |
| ppi | Producer Price Index | NBSC | Index | N | 1 |
| ppi_living | PPI: Living Material | NBSC | Index | Y | 1 |
| ppi_living_food | PPI: Living Material: Food | NBSC | Index | Y | 1 |
| ppi_light | PPI: Light Industry: Annual Rate | NBSC | % | N | 1 |
| gold | Gold Reserve | Wind Database | ten thousand ouce | N | 1 |
| interbank_rate | Interbank Offered Rate | PBC | % | N | 1 |
| loan_rate | Loan Rate (Six Month to One Year) | PBC | % | N | 1 |
| rediscount_rate | Rediscount Rate | PBC | % | N | 1 |
| reserve_ratio | Cash Reserve Ratio | PBC | % | N | 1 |

| Variable | Name | Source | Unit | Seasonally Adjusted | Transformation Applied |
| --- | --- | --- | --- | --- | --- |
| m0 | M0 | PBC | hundred of million RMB | Y | 2 |
| m1 | M1 | PBC | hundred of million RMB | Y | 2 |
| m2 | M2 | PBC | hundred of million RMB | Y | 2 |
| consumption | Total Retail of Consumer Goods | NBSC | hundred of million RMB | Y | 2 |
| public_exp | Public Finance Expenditure | NBSC | hundred of million RMB | Y | 2 |
| indu_add_rate | Industry Value Added: Annual Rate | NBSC | % | Y | 1 |
| power | Power | NBSC | hundred of million kWh | Y | 2 |
| export | Value of Export | GACC | hundred of million dollars | Y | 2 |
| import | Value of Import | GACC | hundred of million dollars | Y | 2 |
| exchange_rate | Foreign Exchange Rate | PBC | RMB per U.S.$ | N | 3 |
| peratio | P/E Ratio: Shanghai Stock Exchange | SSE | % | N | 2 |
| loan | Deposit Balance: Financial Institution | PBC | hundred of million RMB | Y | 2 |
| deposit | Loan Balance: Financial Institution | PBC | hundred of million RMB | Y | 2 |
| volume | Total Futures Trading Volume | CSRC | | Y | 2 |

The transformations are defined as: 1: $(1-B)Z_t$. 2: $(1-B)\log(Z_t)$. 3: $(1-B)^2\log(Z_t)$. For the column of "Seasonal Adjusted", "Y" means the covariate is seasonal adjusted, "N" means not seasonal adjusted.

NBSC: National Bureau of Statistics of China, http://data.stats.gov.cn/.

PBC: People's Bank of China, http://www.pbc.gov.cn/.

GACC: General Administration of Customs of the People's Republic of China, http://www.customs.gov.cn/customs/index/index.html. SSE: Shanghai Stock Exchange, http://english.sse.com.cn/.

CSRC: China Securities Regulatory Commission, http://www.csrc.gov.cn/pub/csrc_en/.

Wind Database: http://www.wind.com.cn/en/edb.html.

Table A.3: covariates loadings in Factor Model (18) for CPI-CN

| Covariate | Factor1 | Factor2 | Factor3 | Factor4 | Factor5 |
|---|---|---|---|---|---|
| ppi_life_food | 0.37 | 0.01 | 0.00 | -0.15 | -0.09 |
| ppi_life | 0.37 | 0.00 | 0.09 | -0.08 | -0.05 |
| ppi | 0.33 | -0.05 | 0.11 | 0.14 | 0.08 |
| rpi | 0.33 | 0.06 | -0.18 | -0.09 | -0.23 |
| ppi_light_rate | 0.26 | 0.08 | 0.06 | 0.06 | 0.16 |
| reserve_ratio | 0.26 | 0.03 | 0.23 | -0.07 | 0.04 |
| rediscount_rate | 0.26 | -0.12 | 0.20 | 0.12 | 0.08 |
| loan_rate | 0.25 | -0.15 | 0.06 | 0.10 | 0.03 |
| cpi_food_meat | 0.24 | 0.16 | -0.22 | -0.35 | 0.13 |
| cpi_food_grain | 0.22 | -0.04 | 0.12 | -0.08 | -0.34 |
| cpi_food_pork | 0.22 | 0.16 | -0.25 | -0.33 | 0.18 |
| import | 0.17 | -0.22 | -0.14 | 0.42 | 0.05 |
| export | 0.16 | -0.12 | -0.15 | 0.41 | 0.13 |
| power | 0.08 | -0.08 | -0.19 | 0.07 | -0.14 |
| exchange_rate | 0.08 | -0.18 | 0.00 | 0.09 | 0.29 |
| tobacco | 0.07 | -0.03 | 0.16 | -0.02 | -0.28 |
| m0 | 0.06 | -0.05 | -0.09 | 0.28 | -0.10 |
| loan | -0.05 | -0.44 | -0.09 | -0.19 | -0.15 |
| interbank_rate | 0.04 | 0.02 | 0.27 | 0.07 | 0.07 |
| indu_add_ratefinal | 0.04 | -0.03 | -0.20 | 0.05 | -0.11 |
| deposit | -0.04 | -0.36 | -0.03 | -0.24 | 0.08 |
| cpi_food_vege | 0.04 | 0.08 | -0.39 | -0.00 | -0.20 |
| gold | -0.03 | -0.03 | -0.09 | -0.00 | 0.11 |
| m2 | -0.03 | -0.46 | -0.08 | -0.15 | -0.12 |
| public_exp | 0.03 | 0.01 | -0.33 | 0.17 | 0.08 |
| consumption | 0.03 | -0.07 | 0.16 | -0.13 | 0.40 |
| cpi_food_fruit | -0.02 | -0.11 | 0.25 | 0.07 | -0.31 |
| rpi_food | 0.02 | 0.22 | -0.18 | 0.04 | -0.19 |
| volumn | 0.01 | -0.15 | -0.30 | 0.03 | 0.23 |
| peratio | -0.01 | -0.12 | -0.06 | 0.04 | -0.22 |
| m1 | 0.01 | -0.39 | -0.02 | -0.22 | 0.08 |
| Proportion in Variance | 0.17733 | 0.087 | 0.071 | 0.070 | 0.048 |
| Cumulative Variance Proportion | 0.177 | 0.264 | 0.335 | 0.405 | 0.453 |

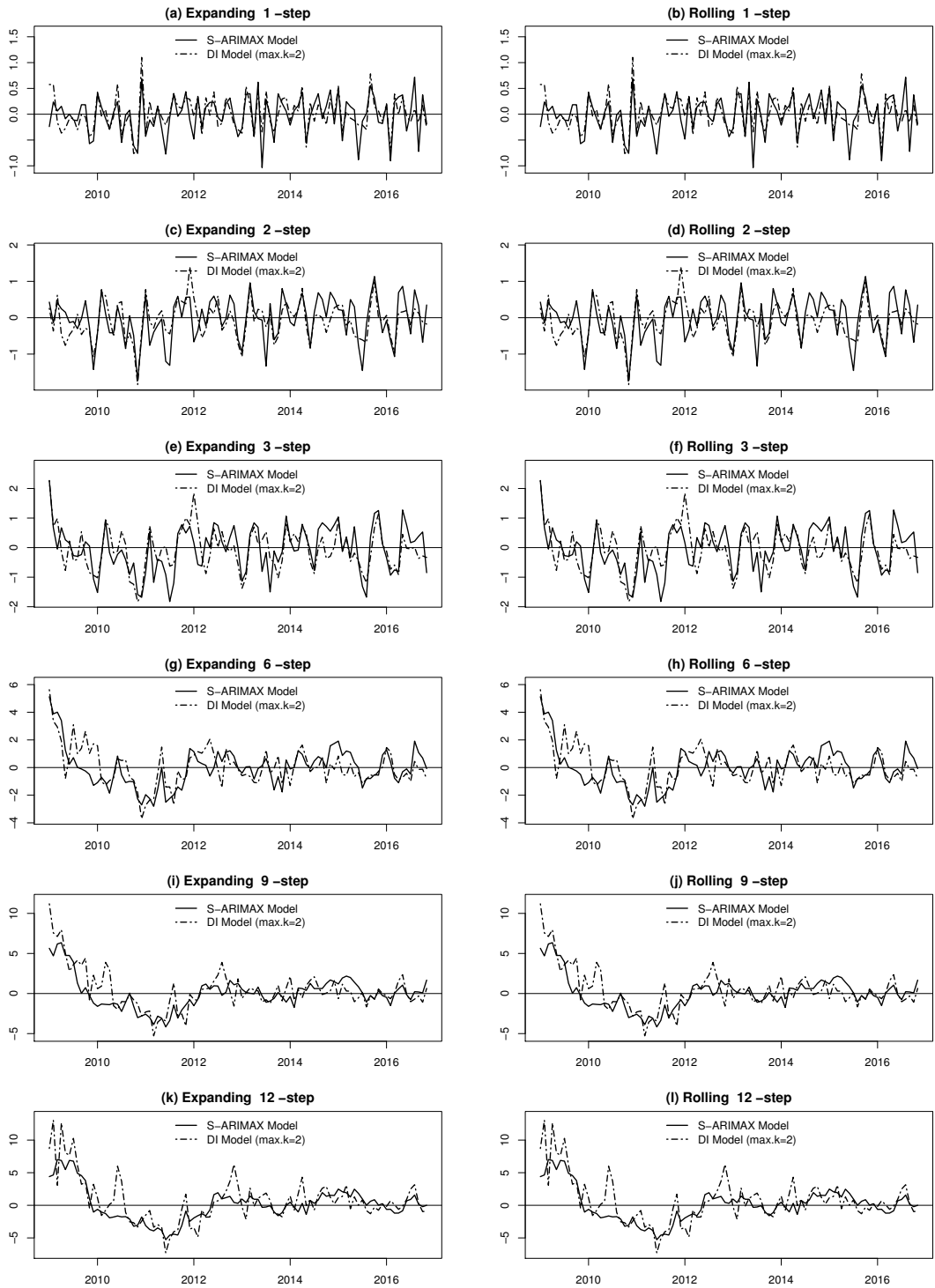

Figure A.1: Out-of-sample forecasting errors of S-ARIMAX model (15) versus the DI approach for the CPI-CN from January 2009 to November 2016 with the expanding and rolling schemes at forecasting horizons ranging from one month to 12 months.

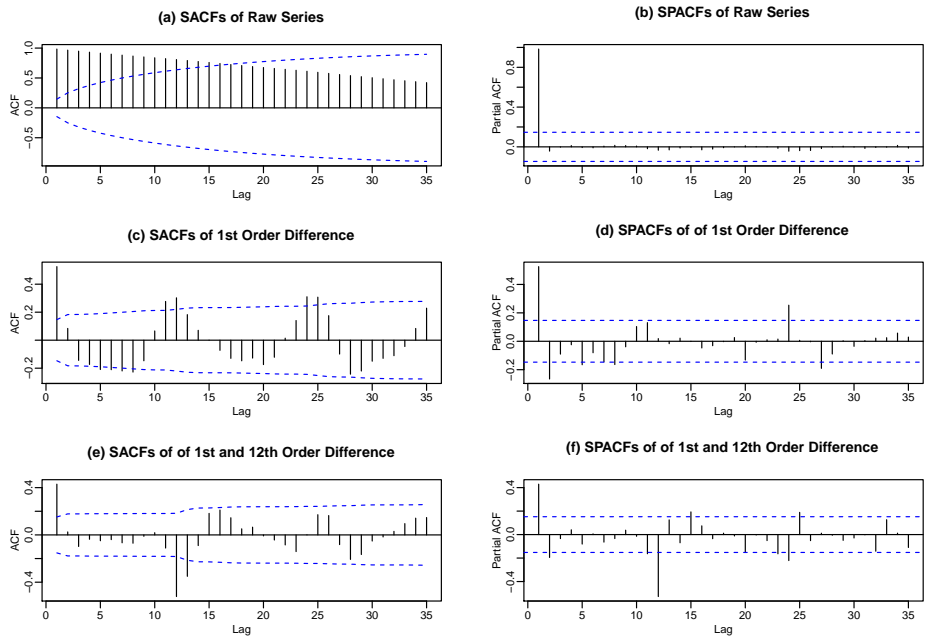

Figure A.2: Sample autocorrelation functions (SACF) and sample partial autocorrelation functions (SPACF) for CPI-US series. Panel (a) and (b): SACF and SPACF of the raw series; Panels (c) an (d): SACF and SPACF of the first differenced series; Panels (e) an (f): SACF and SPACF of the first and the twelfth differenced series.

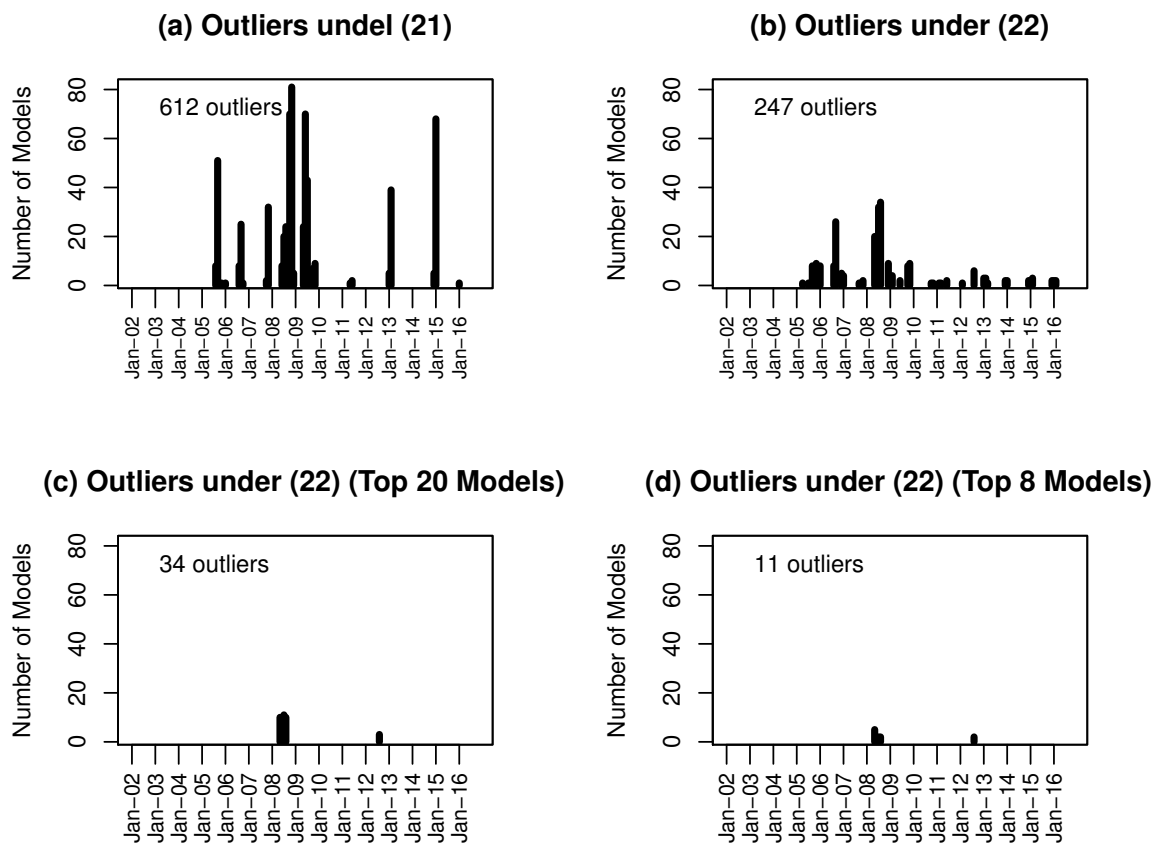

Figure A.3: Monthly Counts of the 81 models that report outliers on each month between January 2002 to November 2016 for CPI-US. Panel (a) is under Model (21), i.e. before considering outliers ; Panel (b) is under equation Model (22), i.e. after considering outliers. Panel (c) and Panel (d) respectively give the results for top 20 models and top 8 models under (22).

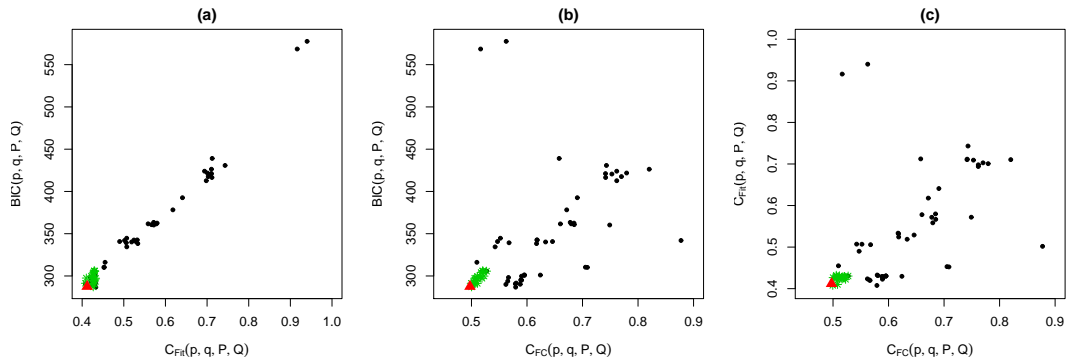

Figure A.4: Pairwise scatter plots of the three model selection criteria: $C_{Fit}(p,q,P,Q)$, $\mathrm{BIC}(p,q,P,Q)$ and $C_{FC}(p,q,P,Q)$ for CPI-US based on the 81 candidate models after considering outliers (Model 22). The red triangle indicates the optimal model with the smallest sum of ranks, and the green stars indicate top 20 models.

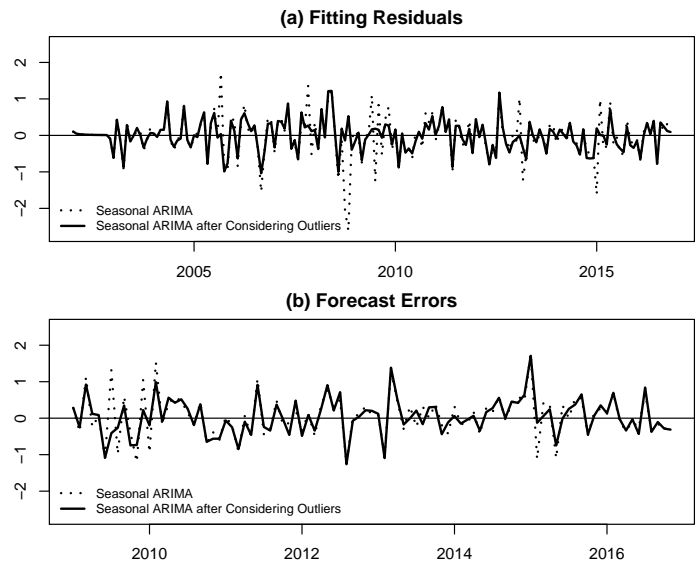

Figure A.5: Fitting Residuals(Panel (a)) and Forecast Errors (Panel (b)) for CPI-US based on the S-ARIMA Model (21) and the Model (23) after considering the outliers.

Table A.4: Forecasting RMSE/$\hat{\sigma}_0$ of the top eight models for CPI-US with forecasting horizon ranging between one month ($h = 1$) and a year ($h = 12$). The smallest one is shown in bold.

| | Expanding Window | | | | | |
|---|---|---|---|---|---|---|
| Forecast Horizon | h=1 | h=2 | h=3 | h=6 | h=9 | h=12 |
| $(2,1,2) \times (0,1,1)$ | **0.491** | 0.866 | 1.267 | 2.346 | 3.023 | 3.274 |
| $(2,1,2) \times (1,1,1)$ | 0.493 | 0.921 | 1.321 | 2.299 | 2.934 | **3.195** |
| $(2,1,0) \times (0,1,1)$ | 0.492 | 0.928 | 1.381 | 2.375 | 3.023 | 3.301 |
| $(2,1,0) \times (1,1,1)$ | 0.498 | 0.941 | 1.386 | 2.371 | 3.048 | 3.322 |
| $(2,1,0) \times (0,1,2)$ | 0.503 | 0.941 | 1.432 | 2.391 | 3.092 | 3.378 |
| $(2,1,2) \times (1,1,2)$ | 0.501 | 0.954 | 1.371 | 2.419 | 3.173 | 3.429 |
| $(0,1,1) \times (1,1,1)$ | 0.500 | **0.861** | **1.182** | **2.256** | **3.006** | 3.280 |
| $(2,1,2) \times (0,1,2)$ | 0.495 | 0.928 | 1.331 | 2.384 | 3.078 | 3.352 |

| | Rolling Window | | | | | |
|---|---|---|---|---|---|---|
| Forecast Horizon | h=1 | h=2 | h=3 | h=6 | h=9 | h=12 |
| $(2,1,2) \times (0,1,1)$ | 0.548 | 1.000 | 1.438 | 2.469 | **3.024** | **3.171** |
| $(2,1,2) \times (1,1,1)$ | 0.555 | 1.034 | 1.467 | 2.529 | 3.143 | 3.297 |
| $(2,1,0) \times (0,1,1)$ | 0.548 | 1.032 | 1.525 | 2.496 | 3.116 | 3.310 |
| $(2,1,0) \times (1,1,1)$ | 0.555 | 1.048 | 1.536 | 2.512 | 3.155 | 3.344 |
| $(2,1,0) \times (0,1,2)$ | 0.556 | 1.057 | 1.541 | 2.518 | 3.197 | 3.379 |
| $(2,1,2) \times (1,1,2)$ | **0.536** | 1.094 | 1.634 | 2.778 | 3.484 | 3.598 |
| $(0,1,1) \times (1,1,1)$ | 0.556 | **0.968** | **1.328** | **2.406** | 3.105 | 3.292 |
| $(2,1,2) \times (0,1,2)$ | 0.570 | 1.080 | 1.532 | 2.523 | 3.130 | 3.285 |

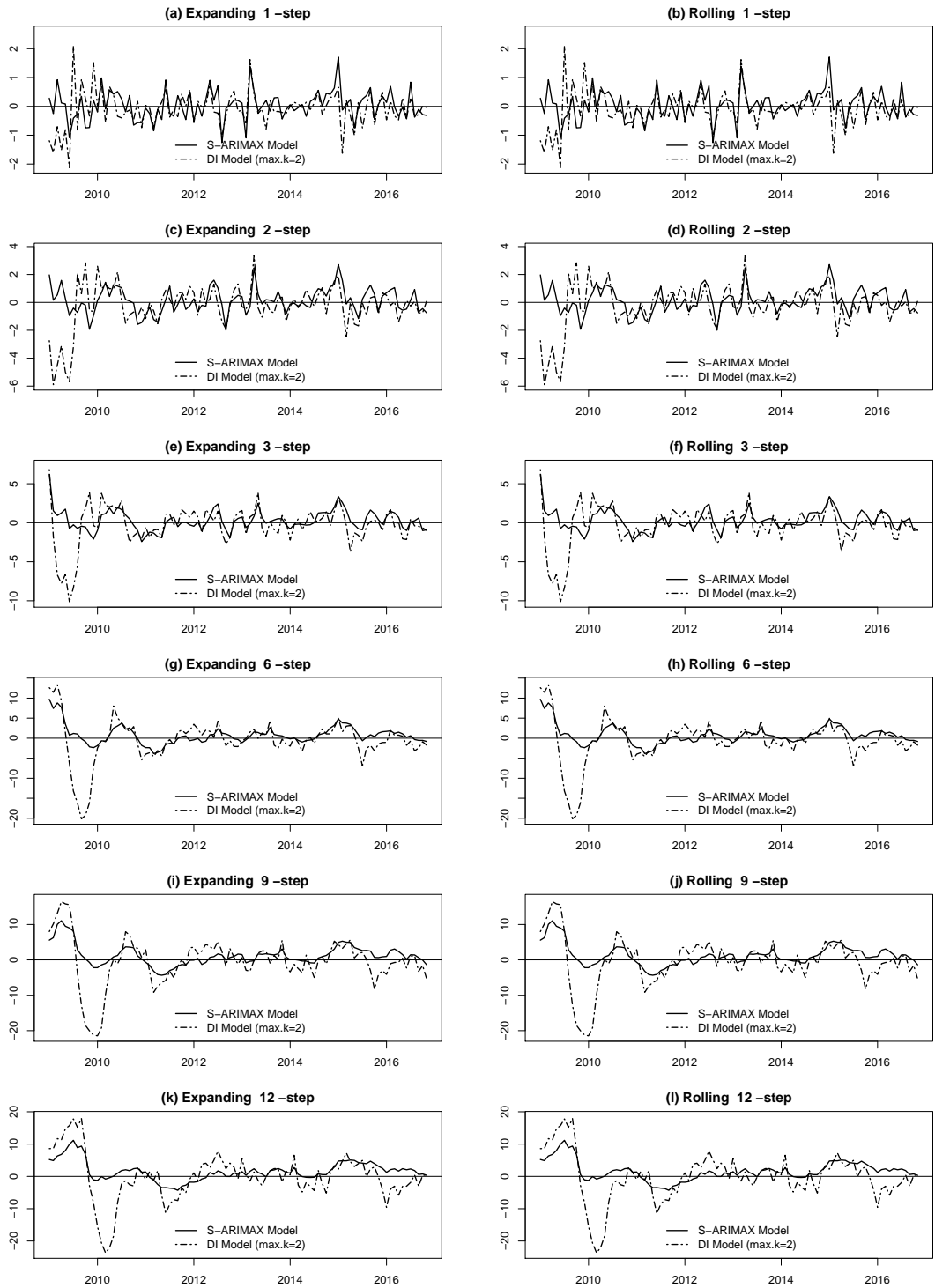

Figure A.6: Out-of-sample forecasting errors of S-ARIMAX model (23) versus the DI approach for the CPI-US from January 2009 to November 2016 with the expanding and rolling schemes at forecasting horizons ranging from one month to 12 months.